\documentclass[aps,prl,superscriptaddress,twocolumn,letterpaper,showpacs,reprint]{revtex4-1}

\usepackage{graphicx}

\begin{document}
\title{Low temperature ballistic spin transport in the $S$=1/2 antiferromagnetic Heisenberg chain compound SrCuO$_2$}

\author{H.~Maeter} \email{h.maeter@physik.tu-dresden.de} \affiliation{Institute for Solid State Physics, TU Dresden, D-01069 Dresden, Germany}

\author{A. A.~Zvyagin} \affiliation{Institute for Solid State Physics, TU Dresden, D-01069 Dresden, Germany}

\affiliation{Institute for Low Temperature Physics and Engineering of the NAS of Ukraine, Kharkov, 61103, Ukraine}

\author{H.~Luetkens} \affiliation{Laboratory for Muon-Spin Spectroscopy, Paul Scherrer Institut, CH-5232 Villigen PSI, Switzerland}

\author{G.~Pascua} \affiliation{Laboratory for Muon-Spin Spectroscopy, Paul Scherrer Institut, CH-5232 Villigen PSI, Switzerland}

\author{Z.~Shermadini} \affiliation{Laboratory for Muon-Spin Spectroscopy, Paul Scherrer Institut, CH-5232 Villigen PSI, Switzerland}

\author{R.~Saint-Martin} \affiliation{Laboratoire de Physico-Chimie de L'Etat Solide, ICMMO, UMR 8182, Universit\'e Paris-Sud, 91405 Orsay, France}

\author{A.~Revcolevschi} \affiliation{Laboratoire de Physico-Chimie de L'Etat Solide, ICMMO, UMR 8182, Universit\'e Paris-Sud, 91405 Orsay, France}

\author{C.~Hess} \affiliation{IFW-Dresden, Institute for Solid State Research, P.O. Box 270116, D-01171 Dresden, Germany}

\author{B.~B\"uchner} \affiliation{IFW-Dresden, Institute for Solid State Research, P.O. Box 270116, D-01171 Dresden, Germany}

\author{H.-H.~Klauss} \affiliation{Institute for Solid State Physics, TU Dresden, D-01069 Dresden, Germany}

\begin{abstract}
We report zero and longitudinal magnetic field muon spin relaxation ($\mu$SR) measurements of the spin $S$=1/2 antiferromagnetic Heisenberg chain material SrCuO$_2$. We find that in a weak applied magnetic field $B_0$ the spin-lattice relaxation rate $\lambda$ follows a power law $\lambda$$\propto$$B_0^{-n}$ with $n$=0.9(3). This result is temperature independent for 5~K$\leq$$T$$\leq$300~K. Within conformal field theory and using the M\"uller ansatz we conclude \textit{ballistic} spin transport in SrCuO$_2$.
\end{abstract}
\pacs{75.10.Pq, 76.75.+i, 73.23.Ad, 76.60.Es}
\maketitle

The copper oxide based low-dimensional electronic systems allow a detailed study of many cooperative quantum phenomena and concepts like Mott insulators, spin-charge separation, quantum phase transitions, and unconventional superconductivity \cite{Zaliznyak04,imada98,*maekwa01,*sachdev11,*lee06}. Quasi 1-d arrangements of corner-sharing CuO$_2$ squares are model compounds for the $S$=1/2 antiferromagnetic Heisenberg chain (AFHC). In these systems charge degrees of freedom are quenched at low energies by strong Coulomb interaction. Spin degrees of freedom are governed by the Heisenberg Hamiltonian $H$=$J\sum_i S_i S_{i+1}$ where $i$ indexes the spins along the chain and the exchange constant $J$ controls the interaction strength between neighboring spins. A model compound for the isotropic $S$=1/2 AFHC is SrCuO$_2$. It is regarded as an almost ideal 1-d system with $J$$\approx$2100~K between neighboring spins of magnetic Cu$^{2+}$ ions \cite{MEU}. In relation to $J$ the residual interchain interaction that causes magnetic order at temperatures below $\approx$2~K \cite{Matsuda97} is very small. 

The ground state (GS) of the $S$=1/2 AFHC is a singlet ($S$=0) state with a continuum of $S$=1/2 excitations, spinons. Recently, using numerical solutions of the Bethe ansatz equations it was shown that the ground state dynamic correlation functions are determined mostly by the 2-spinon continuum \cite{Caux,Muller1,*Muller2,*Muller3}. While its thermodynamic properties have been studied thoroughly, both theoretically and experimentally, the dynamic properties, such as the spin transport have been much less studied: For a strongly interacting system with a non equidistant (due to interactions) spectrum of eigenstates it is an arduous task. As a result theoretical studies are often controversially discussed and even diffusive spin transport has been predicted for some temperature and field regimes \cite{znidaric11,prosen,*prosen1,GB1,*GB2,SPA,*SPA1}.

Recent heat transport experiments revealed an unexpectedly large \textit{magnetic} contribution to the total heat conductivity of $S$=1/2 AFHC materials such as SrCuO$_2$, attributed to the magnetic excitations of the spin chains at low temperatures \cite{hlubek2010,hess07,*sologubenko01}. It is still unclear how this contribution can be understood microscopically. A deeper understanding of the dynamic properties of the prototype Hamiltonian for the $S$=1/2 AFHC, the Heisenberg Hamiltonian $H$=$J\sum_i S_i S_{i+1}$ is essential for progress in this field. Its most important property is integrability, i.e. the existence of infinitely many \textit{local} constants of motion. In general, integrability implies ballistic (spin) transport \cite{zotos96}. In particular, ballistic spin transport is predicted for finite magnetic fields and the ground state \cite{SPA,*SPA1}. However, experimental proof for ballistic spin transport in the $S$=1/2 AFHC is still lacking. Several experimental studies reveal diffusive spin transport in different model compounds of the $S$=1/2 AFHC \cite{pratt06,thurber2001,takigawa96}. In this Letter we present experimental evidence for \textit{ballistic} spin transport in SrCuO$_2$ at low temperature $T$$\ll$$J$ and finite fields $H$$\ll$$J$ based on $\mu$SR experiments. 

The structure of SrCuO$_2$ contains chains running along the crystallographic c axis build of corner sharing CuO squares reminiscent of the CuO layers of cuprate superconductors. In SrCuO$_2$ two chains are joined by sharing their edges, forming a zig-zag chain. The antiferromagnetic coupling between nearest neighbors (NN) is large compared to the ferromagnetic exchange coupling $J'$$\approx$220~K between diagonal Cu spins \cite{RGS}. Hence, it can also be described as a $S$=1/2 chain with ferromagnetic NN and antiferromagnetic next nearest neighbor (NNN) interactions. The resulting weak frustration can produce essential features in the behavior of the spin chain. Inelastic neutron scattering data (INS), however, indicates that both chains are decoupled and show no features associated with frustration \cite{Zaliznyak04}. Neutron diffraction experiments revealed anisotropic spin freezing below $\approx$5~K in SrCuO$_2$ \cite{Zaliznyak99}. This frozen state is also detected by $\mu$SR, but only below $\approx$2~K \cite{Matsuda97}. Thurber and coworkers studied the dynamics of the $q$=0 modes in SrCuO$_2$ by $^{17}$O nuclear magnetic resonance (NMR) \cite{thurber2001}. They find a magnetic field $B_0$ dependence of the spin lattice relaxation rate $1/T_1$$\propto$$B_0^{-n}$ that is consistent with diffusive spin transport, i.e. with $n$=0.5 \cite{thurber2001}.

In this study we used a single crystal of SrCuO$_2$ grown by the traveling solvent floating zone technique from high purity (99.99\%) precursors. The thermal transport properties of this sample have been reported by N.~Hlubek and coworkers in Ref.~\cite{hlubek2010}. A single crystal was oriented and mounted with the crystallographic b axis along the muon beam direction. $\mu$SR experiments were conducted using a $^4$He flow cryostat at the GPS instrument of the Paul Scherrer Institut, Switzerland.
\begin{figure}[tbp]
\includegraphics[width=\columnwidth]{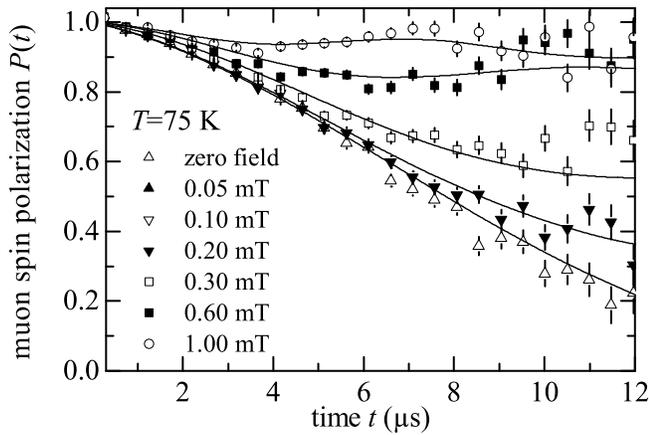}
\caption[]{Typical $\mu$SR time spectra of SrCuO$_2$ at a temperature of 75~K measured for different longitudinal magnetic fields. Solid lines are best fits to the data (see text).} \label{img.spectra}
\end{figure}
In a $\mu$SR experiment, nearly 100\% spin polarized muons are implanted into the sample one at a time. In cuprate materials the positively charged $\mu^+$ usually form a $\approx$1~\AA~long bond with a oxygen ion where they act as magnetic microprobes \cite{muonsite1,*muonsite2,*muonsite3}. In a \textit{non}-magnetic material the muon spins will dephase, i.e., the 100\% initial spin--polarization $P(t)$ will decay as a function of time $t$ due to the random orientation of \textit{nuclear} magnetic dipole fields. The nuclear dipole field distribution can be modeled by an isotropic Gaussian distribution with width $\Delta$. This is well known and can be described by the so called Kubo-Toyabe function $G(t,\Delta,B_0)$, which, in a longitudinal magnetic field $B_0$ has been described by Hayano \cite{hayano1979}. Typically, $B_0$$\approx$5 to 10~mT will decouple the muon spin from the nuclear dipole field distribution.

The quantum spin fluctuations of the $S$=1/2 AFHC cause an additional independent relaxation mechanism for the muon spin ensemble. Rapid fluctuations will cause an exponential relaxation of the muon spin polarization $P(t)$$\propto$$\exp(-\lambda t)$ with the spin lattice relaxation rate $\lambda$. The overall relaxation function is then the product $P(t)$=´$G(t,\Delta,B_0) \exp (-\lambda t)$. In our experiments, $\lambda$ is of the order of 0.01~$\mu$s$^{-1}$ and we study its field and temperature dependence. No signature of muon diffusion has been found, hence $\Delta$=0.085(2)~$\mu$s$^{-1}$ is temperature and field independent. Typical $\mu$SR time spectra and best fits to the data are shown in Fig.~\ref{img.spectra}. 

\begin{figure}[htbp]
\includegraphics[width=\columnwidth]{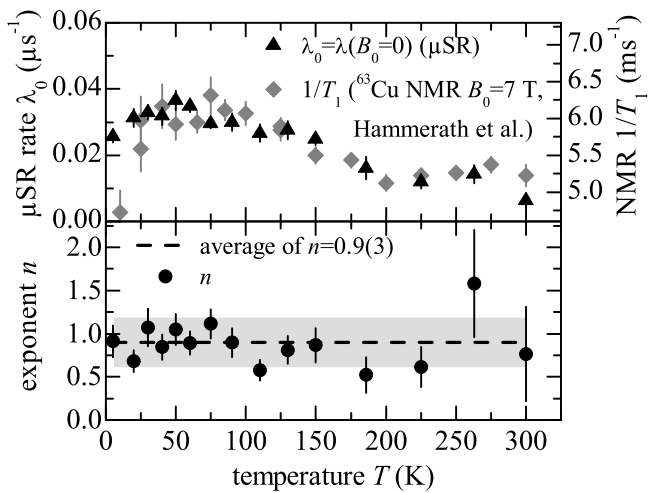} 
\caption[]{\textit{Top}: A comparison of the temperature dependence of $\lambda_0$ with the spin lattice relaxation rate from Ref.~\cite{hammerath10} measured by $^{63}$Cu NMR. The nearly identical temperature dependencies indicate that the muon does not disturb the low energy spin excitation spectrum. \textit{Bottom}: The power law exponent $n$. The shaded area indicates the standard deviation of $n$. All error bars are error estimates from the fits.} \label{img.n}
\end{figure}

To confirm that the measurement is not influenced by the muon as an impurity we compare our data with recent $^{63}$Cu NMR data~\cite{hammerath10} in Fig.~\ref{img.n}. It turns out that both the $\mu$SR and NMR spin-lattice relaxation rates have nearly identical temperature dependencies. This indicates that the muon has little or no influence on the low energy spin excitations of SrCuO$_2$.

For diffusive spin transport in one dimension $\lambda\propto B_0^{-0.5}$ is expected \cite{benner1990,SPA,*SPA1}. Its divergence for $B_0$=0 can be cut-off by 3-d diffusion/coupling or by anisotropy, e.g. dipolar intrachain coupling. A useful description of experimental data has to include this cut-off for low fields/frequencies \cite{mizoguchi1995}:
\begin{equation}
\lambda(B_0)=\lambda_0\left( \frac{1+\sqrt{1+(B_0/2B_{c})^2}}{2(1+(B_0/2B_c)^2)}\right)^n \ . \label{eq.rate2}
\end{equation}
$\lambda_0$=$1/\sqrt{2D_{\parallel}D_{\perp}}$ is the constant value obtained for $B_0$$<$$B_c$ and $\gamma_{\mu} B_c$=$D_{\perp}$, with $n$=0.5. Two diffusion constants $D_{\parallel}$ and $D_{\perp}$ model fast on-chain and slow intrachain diffusion, respectively. We chose this model to maintain comparability with the work of Pratt et al. \cite{pratt06}. Its main purpose is to determine the power law exponent $n$.
\begin{figure}[htbp] 
\includegraphics[width=\columnwidth]{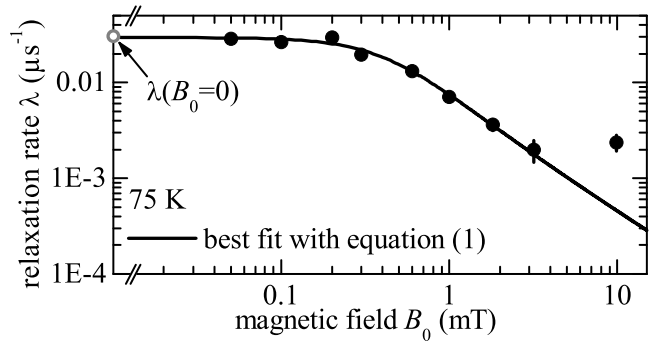}
\caption[]{The field dependence of the relaxation rate $\lambda(B_0)$ shows a clear power law behavior for $B_0$$>$$B_c$$=$0.23(3)~mT and saturates for smaller fields. The solid line is a fit of Eq. (\ref{eq.rate2}) to the data ($B_c$ was optimized simultaneously for all temperatures). The large deviation from the expected field dependence for $B_0$=10~mT is due to the limited time window of the experiment of $\approx$12~$\mu$s. The open symbol indicates $\lambda(B_0$$=$$0)$.} \label{img.rate}
\end{figure}
In experiment \cite{pratt06}, sometimes $n$$\neq$$0.5$ is found and (\ref{eq.rate2}) becomes an empirical model since the derivation in Ref. \cite{mizoguchi1995} is only valid for chains for $n$=0.5 (for a three-dimensional system, where $D_{\perp}$$\ne$0, the equation is formally correct), i.e. in the limit $D_\perp$$\to$0, $\lambda(B_0)$ vanishes or diverges for $n$$>$0.5 or $n$$<$0.5, respectively. We use Eq.~(\ref{eq.rate2}) to analyze our data. For $B_0$$>$$B_c$, $\lambda(B_0)$ follows the power law $B_0^{-n}$, here $B_c$ is called the cut-off field. A typical best fit of Eq.~(\ref{eq.rate2}) to the data is shown in Fig.~\ref{img.rate}. $B_c$=0.23(3)~mT is found to be temperature independent. $n$ also shows no temperature dependence for temperatures between 5 and 300~K, as can be seen in Fig. \ref{img.n}. The average value and its standard deviation are $n$=0.9(3). This result is quantitatively different from the result expected for diffusive spin transport, i.e. $n$=0.5. A microscopic interpretation of this result will be given below. $\lambda_0\approx0.03$~$\mu$s$^{-1}$ could be identified with the inverse of a diffusion constant as in Eq.~(\ref{eq.rate2}) and it shows no anomaly at low temperatures. In fact, all parameters that characterize the spin transport in this system, i.e. $\lambda_0$, $B_c$, and $n$ have no significant temperature dependence for $T\lesssim 150$~K. This indicates that the mechanisms that cause the temperature dependence of the magnetic \textit{heat} transport, which shows a maximum at low temperatures \cite{hlubek2010}, are not relevant for the \textit{spin} transport. This corroborates the conclusions drawn in Ref.~\cite{hlubek2010}, that heat transport of the $S$=1/2 AFHC in SrCuO$_2$ is limited, i.e. rendered diffusive by \textit{extrinsic} perturbations, e.g. phonons.

Next, we will turn to the microscopic interpretation of the experimental results. The relaxation rate for a local probe, i.e. a nuclear, or muon spin in an electron spin system can be expressed as \cite{Moriya1956}
\begin{eqnarray}
\lambda & = & \frac{\gamma_e^2\gamma_N^2}{ 2} \sum_q \bigl(F^z(q) S^{zz}(q,\omega)\bigr.\nonumber \\ 
& & {} \bigl. +\frac{1}{ 4}F^{\perp}(q)[S^{+-}(q,\omega)+ S^{-+}(q,\omega)]\bigr),
\label{T1}
\end{eqnarray}
where $\gamma_{e(N)}$ are the gyromagnetic ratios of electron (nuclear) spins \footnote{We use the units in which $\hbar=k_B=1$.}, $F^{z}(q)$ and $F^{\perp}(q)$ are components of the hyperfine form factors parallel and perpendicular to the external magnetic field, $S^{zz}(q,\omega)$ $S^{+-}(q,\omega)$, and $S^{-+}(q,\omega)$ are the components of the tensor of the dynamic structure factor (DSF) of electron spins, also parallel and perpendicular to the external field. In general, the muon relaxation rate $\lambda$ differs from the nuclear magnetic resonance rate $1/T_1$ only by the form factor. 

A detailed knowledge of the form factor is not required because generally no "filtering" of the DSF by the form factor is expected in cuprate materials. This is due to the low symmetry of the muon site which is usually at a distance of $\approx$1~\AA~from an oxygen ion \cite{muonsite1,*muonsite2,*muonsite3}. 

The experimentally found $\lambda$$\propto$$S^{zz}$$\propto$$1/B_0^n$ with $n$=0.9(3) is in clear disagreement with the model for spin diffusion ($n$=0.5). Note that $n$=0.5 disagrees also with the theoretical work \cite{sachdev94} of Sachdev \footnote{Sachdev \cite{sachdev94} predicts a lifetime of propagating spinons of the order of $\hbar$/$k_BT$, which should be very large at low temperatures.}. This leaves us with two questions: (1) Is the spin transport ballistic?, and (2) How can we understand the observed power law? In the following we will show that the measured power law with $n$$\approx$1 can be understood by the spin excitation spectrum of the ground state of the $S$=1/2 AFHC. In the ground state, exact calculations show that the spin transport is ballistic (see e.g. Ref.~\cite{SPA,*SPA1} and references therein).  Therefore, $n$$\approx$1 proofs ballistic spin transport in the $S$=1/2 AFHC material SrCuO$_2$ because we find $n$=0.9(3) by experiment. One model for the ground state spin excitation spectrum has been given by M\"uller et al. This is a well known model and often used to analyze inelastic neutron scattering data \cite{Zaliznyak04,lake05} In this model $S^{zz}$$\propto$$1/\omega$ in agreement with our experimental result. We will describe this model below.

According to the conjecture of M\"uller et al., the ground state DSF of the $S$=1/2 AFHC is determined by the 2-spinon continuum \cite{Muller1,*Muller2,*Muller3}. In the absence of the magnetic field it has a lower $\epsilon_1(q)$=$\pi J/2 |\sin q|$ and upper bound $\epsilon_2(q)$=$\pi J \sin{q/2}$. The contribution to the ground state DSF of the 2-spinon continuum of the $S$=1/2 AFHC can be written as
\begin{equation}
S_2^{zz}(q,\omega) = A_2 \frac{\Theta(\omega -\epsilon_{1}(q)) \Theta(\epsilon_{2}(q) -\omega) }{ \sqrt{ \omega^2 -\epsilon^2_{1}(q)}} \ , \label{S10}
\end{equation}
where $A_2$ is a constant, and $\Theta(x)$ is the Heaviside step function (See Ref.~\cite{BCK,*KMBFM} for the optimized value of $A_2$).

In the absence of a magnetic field the DSF is isotropic, and at $\omega$$\to$$\omega_0$$\ll$$J$ two points of the reciprocal space, $q$=0 and $q$=$\pi$ contribute mostly to the relaxation rate (\ref{T1}) \cite{Muller1,*Muller2,*Muller3}. On the other hand, a non zero external magnetic field introduces an anisotropy for the components of the dynamic correlation functions of the $S$=1/2 AFHC, shifting the contributing points away from $q$=0, $\pi$. However, in our $\mu$SR experiments the external magnetic field was much smaller than the exchange constant along the spin chain, and this shift is negligible. In $\mu$SR experiments we have $\omega$=$\omega_0$=$\gamma_{\mu}B_0$. In the ballistic regime, according to Eq.~(\ref{S10}) the relaxation rate in the ground state has to be inverse proportional to the value of the external field
\begin{equation}
\lambda \propto \frac{1}{B_0}. \label{rel0}
\end{equation}
Notice that the value $A_2$ is field-dependent \cite{LZ1,*LZ2,*LZ3,*LZ4,*LZ5}. At low fields, this can cause a weak logarithmic dependence of the relaxation rate in addition to the inverse proportionality (\ref{rel0}). The power law (\ref{rel0}), $\lambda$$\propto$$B_0^{-n}$ with $n$=1 valid for the ground state DSF is in agreement with the experimentally found power law with $n$=0.9(3). According to above arguments, this is evidence for ballistic spin transport in SrCuO$_2$. In the following we will discuss the well known effects of exchange anisotropy, temperature and next nearest neighbor (NNN) exchange interaction on the power law (\ref{rel0}).

Uniaxial anisotropy of the exchange interaction can alter the frequency dependence of $S^{zz}(q,\omega)$ and hence the exponent in the power law (\ref{rel0}):
\begin{equation}
S^{zz}(q,\omega) = A \frac{\Theta(\omega -\epsilon_{1}(q)) \Theta(\epsilon_{2}(q) -\omega) }{ (\omega^2 -\epsilon^2_{1}(q))^{\alpha_0} (\epsilon^2_2(q)-\omega^2)^{(1/2)-\alpha_0}} \ .
\label{S20}
\end{equation}
The exponent $\alpha_0$ can be calculated from the finite size corrections to the ground state energy and $\alpha_0$$\equiv$$1- Z^2$=$(\pi -2\eta)/(2\pi -2\eta)$, where $Z$=$\sqrt{\pi/2(\pi -\eta)}$ is the dressed charge of the spin chain model \cite{Zb} in zero external magnetic field. The parameter $\eta$ is related to the magnetic anisotropy of the exchange interaction $\cos \eta$=$J^z/J$ for easy-plane anisotropy, and $\cosh \eta$=$J^z/J$ for easy-axis anisotropy of the spin-spin interaction along the chain ($A$ is $\eta$-dependent). Hence, our experimental results permit a small anisotropy. However, an exact determination of $\eta$ is limited by the experimental accuracy.

In our experiments temperatures were small compared to the intrachain exchange interactions, $T$$\ll$$J,J'$. For low fields we can then estimate the temperature dependence of the relaxation rate $\lambda$ from the temperature dependence of the DSF by conformal field theory \cite{Zb}: 
\begin{equation}
\lambda \propto {\frac{1}{B_0}} \left(\frac{2\pi \alpha T}{v}\right)^{\gamma} \label{rel1}
\end{equation}
where $\alpha$ is the cut-off parameter, $v$=$\pi J\sin (\eta)/2\eta$ is the spinon velocity $v$$\approx$$\pi J/2$ for weak anisotropy, and $\gamma$ is related to the Luttinger liquid exponent, connected to the dressed charge. For the isotropic $S$=1/2 AFHC we have $\gamma$$\approx$0. The temperature dependence of $\lambda_0$ in Fig.~\ref{img.rate} is constant below approx. 150~K hence it is in agreement with $\gamma$$\approx$0.

Formally the double chain in SrCuO$_2$ can be treated by introducing weak NNN exchange interaction. The main effect of the then frustrated NN and NNN exchange interaction is an additional minimum (maximum) for $q$=$\pi/2$ in the lower (upper) boundary of the DSF~\cite{Z06}. However, as we pointed out above, the main contributions to the $\mu$SR (and NMR) spin-lattice relaxation rate have to come from $q$=0, $\pi$, which are not essentially changed by spin frustration in the chain. In addition, it is possible that the low temperature exponent $\gamma$ becomes non zero due to spin frustration \cite{ZK}. This effect should be small because we experimentally find $\gamma$$\approx$0 (see above). Furthermore, any effects of the NNN on the DSF would be detected by inelastic neutron scattering (INS). The absence of any such effect on the DSF measured by INS \cite{Zaliznyak04} shows that up to energies of the order of the exchange interaction $J$$\approx$2100~K the NNN interactions have negligible effects.

It turns out that the results of Monte-Carlo simulations \cite{GB1,*GB2} indicate diffusive behavior of the relaxation rate in the $S$=1/2 AFHC. Those results are valid for high and intermediate temperatures \cite{GB1,*GB2}, i.e. $T$$\gg$$J$, or $T$$\approx$$J$. In our situation we have $T$$\approx$75~K, and $J$$\approx$2000~K, i.e. $T$$\ll$$J$, and we cannot apply the results of Grossjohann et al. \cite{GB1,*GB2} for the explanation of our experiments. Also, similar diffusive behavior was predicted within the field-theoretical approximation \cite{SPA,*SPA1}.  However, the perturbation scheme that Sirker et al. use in Ref.~\cite{SPA,*SPA1} for the calculation of the self-energy (mass operator) cannot be applied in our case, because both perturbation and the main part of the Hamiltonian are determined by the same exchange constant $J$, i.e. formally there is no small parameter in our situation.

It is beyond the scope of this Letter to clarify the differences between our results and previous works \cite{thurber2001,pratt06,takigawa96}. Here we only want to mention the importance of impurities. Hammerath et al. have recently shown \cite{hammerath10} that in SrCuO$_2$ bond disorder leads to the formation of a gap at low temperatures. This is not fully understood but it follows, that in experimental studies that probe the low energy spin excitations, impurities play a crucial role because they can alter the low energy excitation spectrum away from that of the $S$=1/2 AFHC. In most cases this circumstance is not as obvious because no gap is observed. However, it is clear that impurities can change the low energy excitation spectrum of the $S$=1/2 AFHC drastically. The samples that we have studied are of high purity as has been shown by heat transport experiments \cite{hlubek2010}. They demonstrated that the mean free path of the magnetic heat conduction is of the order of 1 $\mu$m \cite{hlubek2010}. This is exceptionally high and can only be due to the high quality of our samples \cite{hlubek2010}. We are therefore lead to believe that the apparent differences between our results and previous results \cite{thurber2001,pratt06,takigawa96} are mostly due to the very low amount of impurities in our samples.

In summary, the magnetic field dependence of the spin lattice relaxation rate $\lambda$ in SrCuO$_2$ is $\lambda$$\propto$$B_0^{-n}$ with $n$=0.9(3). This is in close agreement with $n$=1 which follows from the dynamic structure factor expected for the ground state of the isotropic antiferromagnetic $S$=1/2 Heisenberg chain (AFHC). We conclude that in this system the low energy spin dynamics is determined by the eigenstates of the Heisenberg spin chain Hamiltonian. Therefore, in SrCuO$_2$ spin transport is ballistic at low temperatures and fields $0$$<$$H,T$$\ll$$J$. Furthermore, the absence of a temperature dependence for $T$$\lesssim$150~K of the spin lattice relaxation rate $\lambda(0)$=$\lambda_0$ in zero field indicates that frustration due to NNN interaction do not influence the low energy spin dynamics. In addition it shows that spin and heat transport \cite{hlubek2010} are decoupled in this system. This does not contradict the often cited \textit{magnetic} contribution to the heat conductivity of low dimensional magnets. It follows, that the mechanism causing the temperature dependence of the the magnetic \textit{heat} transport is not relevant for the \textit{spin} transport in SrCuO$_2$.
\begin{acknowledgments}This work was supported by the DFG through Grants No.~HE3439/7 and HE3439/8, by the Mercator Program, and by the European Commission through the NOVMAG Project No.~FP6-032980, the FP6 Contract No. RII3-CT-200 3-505925, and FP7 LOTHERM project (PITN-GA-2009-238475). A.A.Z. thanks the Institute for Chemistry of the V.~Karasin Kharkov National University for support. H.M. thanks F. Hammerath for making her NMR data available for comparison with our $\mu$SR data.\end{acknowledgments}

\end{document}